\documentclass{PoS}

\usepackage{amsmath,amssymb,amsfonts,graphics,graphicx,amscd,amsfonts,epsfig,color}
\usepackage{subfig}
\usepackage{cite}
\usepackage{enumitem}
\usepackage{here}
\newcommand {\beq} {\begin{equation}}
\newcommand {\eeq} {\end{equation}}
\newcommand {\beqa}{\begin{eqnarray}}
\newcommand {\eeqa}{\end{eqnarray}}
\newcommand {\nn} {\nonumber}
\newcommand {\del} {\partial}

\newcommand {\Tr}{\mbox{Tr\,}}

\newcommand {\ee}{\mbox{e}}

\title{New perspectives on the emergence
of (3+1)D expanding space-time
in the Lorentzian type IIB matrix model}
\ShortTitle{New perspectives on the emergence of (3+1)D expanding space-time...}

\author{\speaker{Jun Nishimura}
\\
Theory Center, Institute of Particle and Nuclear Studies,\\
High Energy Accelerator Research Organization (KEK),\\
1-1 Oho, Tsukuba, Ibaraki 305-0801, Japan\\
Department of Particle and Nuclear Physics, 
School of High Energy Accelerator Science,\\
Graduate University for Advanced Studies (SOKENDAI),\\
1-1 Oho, Tsukuba, Ibaraki 305-0801, Japan\\
        E-mail: \email{jnishi@post.kek.jp}}

\abstract{The type IIB matrix model 
is a promising candidate for a nonperturbative formulation 
of superstring theory.
In the Lorentzian version, in particular, 
the emergence of (3+1)D expanding space-time
was observed by Monte Carlo studies of this model.
Here we provide new perspectives on the (3+1)D expanding space-time
that have arised from recent studies.
First it was found that
the matrix configurations generated by the simulation are singular
in that the submatrices representing the expanding 3D space
have only two large eigenvalues associated with the Pauli matrices.
This problem was conjectured to occur due to the approximation used
to avoid the sign problem in simulating the model.
In order to confirm this conjecture,
the complex Langevin method was applied to overcome the sign problem
instead of using the approximation.
The results indeed showed a clear departure from the Pauli-matrix structure,
while the (3+1)D expanding behavior remained unaltered.
It was also found that classical solutions 
obtained
within a certain ansatz show quite generically
a (3+1)D expanding behavior with smooth space-time structure.
}

\FullConference{Corfu Summer Institute 2019 "School and Workshops on Elementary Particle Physics and Gravity" (CORFU2019)\\
		31 August - 25 September 2019\\
		Corfu, Greece}

\begin{document}

%%%%%%%%%%%%%%%%%%%%%%%%%%%%%%%%%%%%%%%%%%%%%%%%%%%%%%%%%%%%%%%%%%%%%%
%%%%%%%%%%%%%%%%%%%%%%%%%%%%%%%%%%%%%%%%%%%%%%%%%%%%%%%%%%%%%%%%%%%%%%
%%%%%%%%%%%%%%%%%%%%%%%%%%%%%%%%%%%%%%%%%%%%%%%%%%%%%%%%%%%%%%%%%%%%%%
\section{Introduction}
%\hspace{0.51cm}
%%%%%%%%%%%%%%%%%%%%%%%%%%%%%%%%%%%%%%%%%%%%%%%%%%%%%%%%%%%%%%%%%%%%%%
%%%%%%%%%%%%%%%%%%%%%%%%%%%%%%%%%%%%%%%%%%%%%%%%%%%%%%%%%%%%%%%%%%%%%%
%%%%%%%%%%%%%%%%%%%%%%%%%%%%%%%%%%%%%%%%%%%%%%%%%%%%%%%%%%%%%%%%%%%%%%

%Phase diagram of QCD

It is widely accepted that 
nonperturbative studies are crucial
in understanding the dynamics of superstring theory.
We consider that this statement is manifested by
the emergence of (3+1)D expanding space-time
in the Lorentzian version of the type IIB matrix model \cite{Kim:2011cr}.
This model was conjectured to be a nonperturbative formulation of
superstring theory \cite{Ishibashi:1996xs},
which is analogous to the lattice gauge theory in QCD.
The model has ten bosonic $N\times N$ Hermitian matrices, which 
are expected to describe ten-dimensional space-time in the large-$N$ limit.
The eigenvalue distribution of the ten bosonic
matrices can collapse to a lower-dimensional manifold, which
may represent
the actual space-time dynamically generated in this model.
%Note, in particular, that this 
In order for this to happen,
the (9+1)D
Lorentz symmetry of the model has to be spontaneously broken.
%% In order for this to happen, 
%% the (9+1)-dimensional
%% Lorentz symmetry of the model has to be spontaneously broken.
%%
%% The aforementioned behavior of the Lorentzian version
%% has been demonstrated by Monte Carlo simulations using a certain
%% approximation.

There are various pieces of evidence for the conjecture
that the type IIB matrix model
is indeed a nonperturbative formulation of
superstring theory.
%The connection to superstring theory can be seen in various ways.
A direct connection to perturbative formulation
of superstring theory can be seen by considering the type IIB superstring 
theory in 10D.
First, the action of the model
%(\ref{S_b}) and (\ref{S_f})
can be regarded as a matrix regularization of
the worldsheet action of type IIB superstring theory
in the Schild gauge \cite{Ishibashi:1996xs}.
This does not imply that the matrix model is merely a formulation
for the ``first quantization'' of superstrings
because multiple worldsheets appear naturally
in the matrix model as block-diagonal configurations,
where each block represents the embedding of a single worldsheet
into the 10-dimensional target space.
Second, under a few reasonable assumptions,
the string field Hamiltonian for type IIB superstring theory
can be derived from Schwinger-Dyson equations for the Wilson loop
operators, which are identified as creation and annihilation operators
of strings \cite{Fukuma:1997en}.
This implies that the type IIB matrix model can reproduce
perturbative expansions in type IIB superstring theory to all orders.

%% Secondly, D-branes in type IIB superstring theory can be
%% described in the matrix model, and the interaction between
%% them can be correctly reproduced \cite{Ishibashi:1996xs}.
%% Thirdly, 

In these connections to type IIB superstring theory, 
the target space coordinates are identified with the eigenvalues
of the matrices $A_\mu$.
In particular, this identification is consistent with
the supersymmetry algebra of the model, in which the translation
that appears from the anti-commutator of supersymmetry generators
is identified with the shift symmetry
%of the bosonic matrices given by 
$A_\mu \mapsto A_\mu + \alpha_\mu {\bf 1}$
of the model, where $\alpha_\mu \in {\bf R}$.
%for $\alpha_\mu \in {\bf R}$.
Also the fact that the model has extended ${\cal N}=2$ supersymmetry
in ten dimensions is consistent with the fact that the model 
actually includes gravity since it is known in field theory
that ${\cal N}=1$ supersymmetry is the maximal one that
can be achieved in ten dimensions without including gravity.

It was also realized that the five types of
superstring theory in ten dimensions
are just different descriptions of the same theory.
Therefore,
it was speculated
that the type IIB matrix model
actually describes the unique underlying theory, although
it takes the form
that has
explicit connection
to perturbative type IIB superstring 
theory \cite{Ishibashi:1996xs,Fukuma:1997en}.

Unfortunately, it is extremely hard to
perform Monte Carlo studies of the type IIB matrix model
due to the so-called sign problem
caused by the complex weight in the partition function.
%both in its Eulicidean and Lorentzian versions.
In the Eulicidean version, it comes from the Pfaffian
that is obtained by integrating out fermionic matrices,
while in the Lorentzian version, it comes from
the phase factor $\ee^{iS_{\rm b}}$ with the bosonic action $S_{\rm b}$.
If we treat the phase of the complex weight by reweighting,
huge cancellation among configurations with different phases occurs,
which makes the calculation impractical.
Recently the complex Langevin 
method (CLM) \cite{Parisi:1984cs,Klauder:1983sp}
has been attracting much attention
as a promising approach to this 
problem \cite{Aarts:2009dg,Aarts:2009uq,Aarts:2011ax,Seiler:2012wz,Nishimura:2015pba,Nagata:2015uga,Nagata:2016vkn,Ito:2016efb,Aarts:2017vrv,Scherzer:2018hid}.
In particular, it was applied
successfully to the Euclidean version of 
the 6D type IIB matrix model \cite{Anagnostopoulos:2017gos},
and the spontaneous breaking of
the rotational SO(6) symmetry to SO(3) suggested 
by the Gaussian expansion method \cite{Aoyama:2010ry} was confirmed.
Recently this work has been extended to the original 
Euclidean 10D type IIB matrix model \cite{Anagnostopoulos:2020xai}.
Similarly to the 6D version,
the rotational SO(10) symmetry 
was found to be spontaneously broken to SO(3) as suggested 
by the Gaussian expansion method \cite{Nishimura:2011xy}.
%was confirmed.

In ref.~\cite{Kim:2011cr} as well as in more recent
work \cite{Ito:2013ywa,Ito:2015mxa,Ito:2017rcr,Azuma:2017dcb}
on the Lorentzian type IIB matrix model and its simplified models,
the sign problem was avoided by
integrating out the scale factor
of the bosonic matrices by hand, which yields a function of
the bosonic action $S_{\rm b}$ sharply peaked at the origin.
Approximating this function by a sharply peaked Gaussian function,
we can perform Monte Carlo simulations without the sign problem.
The emergence of (3+1)D expanding space-time
%, for instance, 
was obtained in this way \cite{Kim:2011cr}.
%\cite{Kim:2011cr} 
The expanding behavior for a longer time was 
investigated by simulating the simplified models.
%of the type IIB matrix model.
The obtained results suggested a scenario for the full model 
that the expansion is exponential
at early times \cite{Ito:2013ywa},
which is reminiscent of the inflation,
and that it turns into a power law \cite{Ito:2015mxa}
at later times, which is reminiscent of
the Friedmann-Robertson-Walker universe in the radiation dominated era.
See also refs.~\cite{Yang:2015vna,Kim:2018mfv,Tomita:2015let}
for closely related work.

It has been found recently, however, that the matrix configurations
generated by the simulation is singular
in that the submatrices representing the expanding 3D space
%are essentially proportional to the Pauli matrices 
have only two large eigenvalues associated with 
the Pauli matrices \cite{Aoki:2019tby}.
This problem has been attributed to the aforementioned
approximation used to avoid the sign problem since 
the function obtained after integrating out the scale factor is
actually complex-valued, and the effect of the phase is 
%neglected
not taken into account.
% in the approximation.
%strong enough to eliminate the contribution from all the configurations
%with nonzero $S_{\rm b}\neq 0$, which is typically a 
%quantity of order $N^2$.
It was realized that the approximation actually amounts
to replacing the phase factor $\ee^{iS_{\rm b}}$ 
by a positive definite weight $\ee^{c S_{\rm b}}$ with 
some constant $c >0$.
This new interpretation of the simulation provides
clear understanding of the observed
Pauli-matrix structure and the (3+1)D expanding behavior.
It has also been argued that
a regular space-time may be obtained if the phase factor $\ee^{iS_{\rm b}}$ 
is used correctly.
This is a very nontrivial issue, however, since losing 
the Pauli-matrix structure may also imply 
losing the (3+1)D expanding behavior at the same time.
A piece of evidence that this is possible has been provided
by generating classical solutions of the Lorentzian type IIB matrix 
model \cite{Hatakeyama:2019jyw}, where it was found that
the solutions obtained within a certain ansatz 
show quite generically a (3+1)D expanding behavior with smooth space-time structure.

In ref.~\cite{Nishimura:2019qal}, which we review in this article,
we addressed this issue by using the CLM
to solve the sign problem
% in the Lorentzian matrix model 
instead of using the aforementioned approximation.
%performing Monte Carlo simulations
%of the model without the sign problem obtained by 
Note that the Lorentzian type IIB matrix model
needs to be regularized in some way or another 
because the phase factor $\ee^{iS_{\rm b}}$
in the partition function cannot suppress
the contribution from the bosonic matrices with arbitrary 
% in some way or another 
large elements.\footnote{This situation is in sharp contrast to the 
Euclidean version \cite{Krauth:1998xh,Austing:2001pk},
in which the phase factor $\ee^{iS_{\rm b}}$ is 
replaced by $\ee^{-S_{\rm b}^{({\rm E})}}$, where 
$S_{\rm b}^{({\rm E})}$ is a real non-negative quantity.}
Here we use the infrared cutoffs on both the spatial and temporal
matrices analogous to the ones used in the previous work \cite{Kim:2011cr}.
We also find it useful to introduce
two deformation parameters $(s,k)$,
%in the action, 
which correspond to the Wick rotations
on the worldsheet and in the target space, respectively.
These parameters enable us to interpolate between the Lorentzian version
$(s,k)=(0,0)$ 
and the Euclidean version $(s,k)=(1,1)$ .

First we focus on
$(s,k)=(-1,0)$ 
in the deformation parameter space, where we do not have 
the sign problem. In fact, this case corresponds to 
the approximate model investigated in our previous work.
We observe the emergence of 
(3+1)D expanding space-time with the Pauli-matrix structure.
Then we tune the worldsheet deformation parameter $s$ 
close to that for the Lorentzian model ($s=0$)
keeping the target space deformation parameter $k$ in such a way that
the space-time noncommutativity is minimized.
%; i.e., $k=(1+s)/2$. 
There, we find it possible to obtain
smoother space-time structure
% different from Pauli matrices,smoother than
without losing the (3+1)D expanding behavior.
The deviation from the Pauli-matrix structure was not seen 
for the matrix size $N \le 64$ within the parameter region 
that can be explored by the CLM, 
and it becomes more prominent as we increase $N$ from 128 to 192.
We consider that the two deformation parameters $s$ and $k$ should
be tuned eventually to $(s,k)=(0,0)$
% for the Lorentzian model 
in the large-$N$ limit.
Whether a smooth classical space-time picture appears
in that limit at sufficiently late time is an important open question,
which can be answered
along the line of this research.

The rest of this article is organized as follows. 
In section \ref{sec:definition} we
define the Lorentzian type IIB matrix model
%in some detail,
%how to define the Lorentzian type IIB matrix model,
and introduce the infrared cutoffs as well as the
two deformation parameters $s$ and $k$.
In section \ref{sec:CLM} we discuss how we apply the CLM
to the Lorentzian type IIB matrix model.
In section \ref{sec:expanding} we focus on
$(s,k)=(-1,0)$ in the deformation parameter space, 
which corresponds to 
the approximate model investigated in the previous work.
Indeed we observe the emergence of 
(3+1)D expanding space-time with the Pauli-matrix structure.
In section \ref{sec:departure} we show our results for
the worldsheet deformation parameter $s$ 
close to that for the Lorentzian model ($s=0$)
with the target space deformation parameter $k$ 
%being fixed to $k=(1+s)/2$.
chosen in such a way that
the space-time noncommutativity is minimized.
We observe a clear departure from the Pauli-matrix structure,
while the (3+1)D expanding behavior is still being observed.
Section \ref{sec:summary} is devoted to a summary and discussions.

\section{Definition of the Lorentzian type IIB matrix model}
\label{sec:definition}

The action of the Lorentzian type IIB matrix 
model
is given by \cite{Ishibashi:1996xs}
\begin{eqnarray}
S & = & S_{{\rm b}}+S_{{\rm f}} \ ,
\label{eq:S_likkt}\\
S_{{\rm b}} & = & 
\frac{1}{4g^{2}}{\rm Tr}
\left(\left[A_{\mu},A_{\nu}\right]
\left[A^{\mu},A^{\nu}\right]\right) \ ,
\label{eq:Sb}\\
S_{{\rm f}} & = & 
-\frac{1}{2g^{2}}{\rm Tr}
\left(\Psi_{\alpha}\left(\mathcal{C}
\Gamma^{\mu}\right)_{\alpha\beta}
\left[A_{\mu},\Psi_{\beta}\right]\right) \ ,
\label{eq:Sf-1}
\end{eqnarray}
where the bosonic variables
$A_{\mu}$ $\left(\mu=0,\ldots,9\right)$
and the fermionic variables
$\Psi_{\alpha}$ $\left(\alpha=1,\ldots,16\right)$
are $N\times N$ Hermitian matrices. 
$\Gamma^{\mu}$ are 10D gamma-matrices
after the Weyl projection and $\mathcal{C}$ is the charge conjugation
matrix. The ``coupling constant'' $g$ is merely a scale parameter
in this model since it can be absorbed by rescaling $A_{\mu}$ and
$\Psi$ appropriately. 
The indices $\mu$ and $\nu$
are contracted using the Lorentzian metric 
$\eta_{\mu\nu}={\rm diag}\left(-1,1,\ldots,1\right)$.
%, you can always obtain the Euclidean version
%of the model.
The Euclidean version can be obtained by making
a ``Wick rotation'' $A_0 = i A_{10}$, where $A_{10}$ 
is Hermitian.

The partition function 
for the Lorentzian version
% of the type IIB matrix model
is proposed in ref.~\cite{Kim:2011cr} as 
\begin{equation}
Z=\int dAd\Psi\, e^{iS}
\label{Z-Likkt1}
\end{equation}
with the action \eqref{eq:S_likkt}. 
The ``$i$'' in front of the action
is motivated from the fact that
the string worldsheet metric should also have 
a Lorentzian signature. 
By integrating out the fermionic matrices,
we obtain the Pfaffian 
\begin{equation}
\int d\Psi\, e^{iS_{{\rm f}}} =
{\rm Pf}\mathcal{M}\left(A\right)  \ ,
\end{equation}
which is real unlike in the Euclidean case \cite{Anagnostopoulos:2013xga}.
%% Then \eqref{Z Likkt1} becomes 
%% \begin{equation}
%% Z=\int dA\,{\rm Pf}\mathcal{M}\left(A\right)\, 
%% e^{iS_{{\rm b}}}\ .
%% \label{Z Likkt2}
%% \end{equation}
Note also that the bosonic action \eqref{eq:Sb}
can be written as 
\begin{eqnarray}
S_{\rm b}  =  
\frac{1}{4g^{2}}{\rm Tr}\left(F_{\mu\nu}F^{\mu\nu}\right)
 =  \frac{1}{4g^{2}}
\left\{ {\rm -2Tr}\left(F_{0i}\right)^{2}+
{\rm Tr}\left(F_{ij}\right)^{2}\right\} \ ,
\label{decomp-Sb}
\end{eqnarray}
where we have introduced
the Hermitian matrices $F_{\mu\nu}=i\left[A_{\mu},A_{\nu}\right]$.
Since the two terms
% in (\ref{decomp-Sb})
in the last expression
have opposite signs,
$S_{{\rm b}}$ is not positive semi-definite,
and it is not bounded from below.

In order to make the partition function \eqref{Z-Likkt1} finite,
we need to introduce infrared cutoffs 
in both the temporal and spatial directions, 
for instance, as
\begin{eqnarray}
\frac{1}{N}{\rm Tr}\left(A_{0}\right)^{2} 
& \le & \kappa
%\frac{1}{N}{\rm Tr} \left(A_{i}\right)^{2} 
L^2
\ ,\label{eq:t_cutoff}\\
\frac{1}{N}{\rm Tr}\left(A_{i}\right)^{2} 
& \le & L^{2} \ .
\label{eq:s_cutoff}
\end{eqnarray}

We can use
the ${\rm SU}\left(N\right)$ symmetry of the model
to bring the temporal matrix $A_{0}$ into the diagonal form
\begin{equation}
A_{0}={\rm diag}\left(\alpha_{1},\ldots,\alpha_{N}\right)\ ,
\quad \quad
{\rm where~} \alpha_{1}<\cdots<\alpha_{N} \ .
\label{eq:diagonal gauge}
\end{equation}
By ``fixing the gauge'' in this way,
we can rewrite the partition function (\ref{Z-Likkt1}) as
%integration over $A_\mu$ as
\beqa
\label{gauge-fixing}
%\int dA \, e^{iS}
Z &=& \int  \prod_{a=1}^{N}d\alpha_{a}\,
\Delta (\alpha)^2  \int dA_i \, e^{iS_{\rm b}} 
{\rm Pf}\mathcal{M}\left(A\right) \ , 
\\
\Delta (\alpha) &\equiv &
\prod_{a>b}^{N}
\left(\alpha_{a}-\alpha_{b}\right) \ ,
\label{A0diag}
\eeqa
where $\Delta(\alpha)$ is the van der Monde determinant.
The factor $\Delta (\alpha)^2$ 
in (\ref{gauge-fixing})
appears from the Fadeev-Popov procedure
for the gauge fixing, and it acts as a repulsive potential 
between the eigenvalues $\alpha_i$ of $A_0$.

%Since the eigenvalues of $A_{0}$ represent the real time coordinate 
%in the Lorentzian model,
We can extract a time-evolution
from configurations generated by simulating (\ref{gauge-fixing}).
%Let us explain how to do it.
A crucial observation is that 
the spatial matrices $A_{i}$ have
a band-diagonal structure
in the SU($N$) basis in which $A_{0}$ 
has the diagonal form (\ref{eq:diagonal gauge}).
More precisely, there exists some integer $n$ such that
the elements of spatial matrices
$\left(A_{i}\right)_{ab}$ for $\left|a-b\right|>n$ are 
much smaller than those for $\left|a-b\right|\leq n$.
Based on this observation,
we may naturally consider $n\times n$ submatrices of $A_i$ defined as
\begin{equation}
\left(\bar{A}_{i}\right)_{IJ}\left(t\right)
\equiv\left(A_{i}\right)_{\nu+I,\nu+J} \ ,
\label{eq:def_abar}
\end{equation}
where $I,J=1,\ldots , n$, $\nu=0,1,\ldots , N-n$,
and $t$ is defined by 
\begin{equation}
t=\frac{1}{n}\sum_{I=1}^{n}\alpha_{\nu+I} \ .
\label{eq:def_t}
\end{equation}
We interpret the $\bar{A}_i(t)$ 
as representing the state of the universe at time $t$.
%in (\ref{eq:def_abar}) 

%corresponding to the $n\times n$ matrices $\bar{A}_{i}$. 
%Using the state $\bar{A}_{i}\left(t\right)$, 

Using $\bar{A}_{i}(t)$,
we can define, for example,
the extent of space at time $t$ as 
\begin{equation}
R^{2}\left(t\right)=
\left\langle \frac{1}{n}{\rm tr}\sum_{i}
\left(\bar{A}_{i}\left(t\right)\right)^{2}\right\rangle \ ,
\label{eq:def_rsq}
\end{equation}
where the symbol ${\rm tr}$ represents
a trace over the $n\times n$ submatrix.
%One can also probe the spatial extentin each direction by considering 
We also define
the ``moment of inertia tensor'' 
\begin{equation}
T_{ij}\left(t\right)
=\frac{1}{n}{\rm tr}
\Big(\bar{A}_{i}(t) \bar{A}_{j}(t)\Big) \ ,
\label{eq:def_tij}
\end{equation}
which is a $9\times9$ real symmetric matrix. 
The eigenvalues of $T_{ij}\left(t\right)$,
which we denote by $\lambda_{i}\left(t\right)$ with the order
\begin{equation}
\lambda_{1}\left(t\right)>\lambda_{2}
\left(t\right)>\cdots>\lambda_{9}\left(t\right)
\end{equation}
represent the spatial extent in each of 
the nine directions at time $t$.
Note that the expectation values 
$\left\langle \lambda_{i}\left(t\right)\right\rangle $
tend to be equal in the large-$N$ limit if the SO(9) symmetry is
not spontaneously broken. 
This is the case at early times of the time-evolution.
After a critical time $t_{{\rm c}}$, on the other hand,
it was found \cite{Kim:2011cr}
that the three largest eigenvalues
$\left\langle \lambda_{i}\left(t\right)\right\rangle$ 
($i=1$, $2$, $3$)
%three of the eigenvalues of $T_{ij}\left(t\right)$
become significantly larger than the rest,
%% \begin{equation}
%% \lambda_{1}\left(t\right)>
%% \lambda_{2}\left(t\right)>
%% \lambda_{3}\left(t\right)\gg
%% \lambda_{4}\left(t\right)>\cdots
%% \:{\rm for}\: t>t_{{\rm c}} \ .
%% \end{equation}
which implies that
the SO(9) symmetry is spontaneously broken down to SO(3).

Here we introduce two deformation parameters $s$ and $k$,
which correspond to Wick rotations on the worldsheet and 
in the target space, respectively.
Let us introduce $\tilde{S}=- i S_{\rm b}$ so that
the factor $e^{iS_{\rm b}}$
in the partition function (\ref{gauge-fixing}) 
is rewritten as $e^{- \tilde{S}}$.
We introduce the first parameter $s$ ($-1 \le s \le 1$)
corresponding to the Wick rotation on the worldsheet as
%by replacing the bosonic action (\ref{decomp-Sb}) with
\begin{equation}
%S = N \beta \ee^{- i \frac{\pi}{2}(1-s)}
\tilde{S} = - i N \beta \, \ee^{ i s \pi/ 2}
\left\{ 
- \frac{1}{2} \Tr (F_{0i})^2 
+ \frac{1}{4} \Tr (F_{ij})^2 
\right\} \ ,
\label{sdef-action}
\end{equation}
where
%the overall factor 
$\beta = \frac{1}{g^2 N}$.
%is introduced for convenience.
The second parameter $k$ ($0 \le k \le 1$)
corresponding to the Wick rotation in the target space
can be introduced by the replacement $A_0 \mapsto  \ee^{- i k \pi /2} A_0$.
The action (\ref{sdef-action}) becomes
\begin{equation}
\tilde{S} = - i N \beta \,
\ee^{ i s \pi/ 2}
% \ee^{- i \frac{\pi}{2}(1-s)}
\left\{ 
- \frac{1}{2} \ee^{- i k \pi} \Tr (F_{0i})^2 
+ \frac{1}{4} \Tr (F_{ij})^2 
\right\}  \ ,
\label{sdef-action2}
\end{equation}
and the ${\rm Pf}\mathcal{M}\left(A\right)$ in
(\ref{gauge-fixing})
should be replaced by 
${\rm Pf}\mathcal{M}(\ee^{- i k \pi /2} A_0, A_i)$.
% in modified analogously.
The Lorentzian model is retrieved at $(s,k)=(0,0)$,
whereas the Euclidean model corresponds to setting $(s,k)=(1,1)$.
%% If we disregard the infrared cutoffs
%% (\ref{eq:t_cutoff}) and (\ref{eq:s_cutoff}),
%% the Euclidean model can be obtained at $(s,k)=(1,1)$.

Note that 
the coefficient of the first term in
(\ref{sdef-action2})
can be made
real non-negative by choosing the parameters so that
$ i \ee^{ i s \pi/ 2} \ee^{- i k \pi} = 1 $,
which implies $k=(1+s)/2$.
% $\ee^{- i \frac{\pi}{2}(1-s)} \ee^{- i k \pi} = -1$, 
For this choice, the bosonic action 
is most effective in
minimizing the noncommutativity between
the spatial matrices $A_i$ and the temporal matrix $A_0$.
For $0 \le k < s/2$, on the other hand, 
the real part of the coefficient becomes negative,
which favors maximum noncommutativity between $A_i$ and $A_0$.
%which causes the eigenvalues of $A_0$ to repel each other.
As a result, the eigenvalues of $A_0$ 
lump up into two clusters separated from each other,
and we cannot obtain a continuous time.
The Lorentzian model $(s,k)=(0,0)$ lies on the boundary of this
unphysical region.
Here we keep away from this region by
restricting ourselves to the cases satisfying $k=(1+s)/2$.

Taking into account the infrared cutoffs 
(\ref{eq:t_cutoff}) and (\ref{eq:s_cutoff}),
we arrive at the partition function
\beqa
Z &=& 
\int  \prod_{a=1}^{N}d\alpha_{a}\,
\Delta (\alpha)^2  \int dA_i \, e^{- \tilde{S}} 
{\rm Pf}\mathcal{M}(\ee^{- i k \pi /2} A_0, A_i) \times 
\nn
\\
&~& 
%Z=\int dAd\Psi\, e^{-S}
%{\rm Pf}\mathcal{M}\left(A\right) 
\theta\left(\kappa L^2 -  \frac{1}{N}{\rm Tr}(A_{0})^{2}\right)
\theta\left(L^2 -  \frac{1}{N}{\rm Tr}(A_{i})^{2} \right)
\ ,
\label{Z-Likkt-n1-cutoff}
\eeqa
where $\theta(x)$ is the Heaviside step function
and $\tilde{S}$ is given by (\ref{sdef-action2}).
By rescaling $A_\mu \mapsto L A_\mu$ 
and $\beta \mapsto L^{-4} A_\mu$,
we can set $L=1$ without loss of generality.

%%%%%%%%%%%%%%%%%%%%%%%%%%%%%%%%%%%%%%%%%%%%%%%%%%%%%%%%%%%%%%%%%%%%%%%%%%%%%%%%%%%%%%%%%%
\section{The complex Langevin method}
% and criterion for correct convergence}
\label{sec:CLM}
%\label{sec:correct convergence}
%%%%%%%%%%%%%%%%%%%%%%%%%%%%%%%%%%%%%%%%%%%%%%%%%%%%%%%%%%%%%%%%%%%%%%%%%%%%%%%%%%%%%%%%%%

In this section, we review the CLM and discuss how we 
apply it to the Lorentzian type IIB matrix model.

\subsection{Brief review of the CLM}
\label{sec:CLM-review}

%the criterion for correct convergence
Let us consider a system
\begin{align}
  Z=\int dx \, w(x) 
\label{original-theory}
\end{align}
of $N$ real variables $x_k$ ($k=1 ,\cdots , N$)
%defined by the partition function
as a simple example. 
Here the weight $w(x)$ is a complex-valued function, 
which causes the sign problem.

%\subsection{complex Langevin method}

%% %Under the complexification, 
%% Accordingly, the weight function $w(x)$ and 
%% the observables $\mathcal O(x)$ are extended holomorphically as
%% $w(x)\to w(z)$ and $\mathcal O(x)\to \mathcal O(z)$.

In the CLM, 
the original real variables $x_k$ 
are complexified as $x_k\to z_k=x_k+iy_k\in \mathbb{C}$
and one considers a fictitious time evolution of
the complexified variables $z_k$ using
the complex Langevin equation
% for $z_k$, which is 
given, in its discretized form, by
\begin{align}
z^{(\eta)}_k(t+\epsilon)=z^{(\eta)}_k(t)
+\epsilon \, v_k(z^{(\eta)}(t))+\sqrt{\epsilon} \, \eta_k(t) \ ,
\label{Langevin}
\end{align}
where $t$ is the fictitious time with a stepsize $\epsilon$.
The second term $v_k(z)$ on the right-hand side is called 
the drift term, which is defined by
holomorphic extension of the one
\begin{align}
v_k(x)=w(x)^{-1}\frac{\partial w(x)}{\partial x_k}
\end{align}
for the real variables $x_k$.
The variables $\eta_k(t)$ appearing on the right-hand side of 
eq.~(\ref{Langevin})
are a real Gaussian noise 
with the probability distribution
$\propto e^{-\frac{1}{4}\sum_t\eta_k(t)^2}$,
which makes the time-evolved variables $z^{(\eta)}_k(t)$ stochastic.
The expectation values
with respect to the noise $\eta_k(t)$ 
are denoted as $\langle \cdots \rangle_\eta$ in what follows.
%$\langle\eta_k(t)\eta_j(t')\rangle=2\delta_{tt'}\delta_{ij}$.

Let us consider the expectation value of 
an observable $\mathcal O(x)$.
In the CLM, one computes the expectation value of the 
holomorphically extended observable $\mathcal O(x+iy)$ as
\begin{align}
%  \Phi(t)= 
\Big\langle \mathcal O(z^{(\eta)}(t)) \Big\rangle_\eta
=\int dx \, dy \, \mathcal O(x+iy) P(x,y;t) \ ,
  \label{Phi}
\end{align}
where $P(x,y;t)$ is the probability distribution of 
$x^{(\eta)}(t)$ and $y^{(\eta)}(t)$ defined by
\begin{align}
  P(x,y;t)=\Big \langle 
\delta(x-x^{(\eta)}(t))\delta(y-y^{(\eta)}(t)) \Big\rangle_\eta \ .
  \label{P}
\end{align}
Then, the correct convergence of the CLM implies the equality
\begin{align}
  \lim_{t\to \infty}\lim_{\epsilon \to 0}
\Big\langle \mathcal O(z^{(\eta)}(t)) \Big\rangle_\eta
%\Phi(t)
=\frac{1}{Z}\int dx \, \mathcal O (x)w(x) \ ,
  \label{key}
\end{align}
where the right-hand side is the expectation value 
of $\mathcal O(x)$ in the original theory (\ref{original-theory}).
A proof of eq.~\eqref{key} was given
in refs.~\cite{Aarts:2009uq,Aarts:2011ax}, where the notion of
the time-evolved observable $\mathcal O (z;t)$ plays a crucial role.
In particular, it was pointed out that the integration by parts 
used in the argument cannot be justified 
when the probability distribution \eqref{P} falls off slowly
in the imaginary direction.
In ref.~\cite{Nishimura:2015pba},
it was noticed that the wrong convergence
associated with the zeroes of the fermion determinant
\cite{Mollgaard:2013qra}
is actually due to 
the slow fall-off of the probability distribution \eqref{P} 
toward the singularities of the drift term.

While this argument provided theoretical understanding of the cases
in which the CLM gives wrong results, the precise condition on the 
probability distribution was not specified.
Furthermore, there is actually a subtlety
in defining the time-evolved observable.
Recently ref.~\cite{Nagata:2016vkn} provided
a refined argument for justification
of the CLM \cite{Nagata:2016vkn}, which showed that
the probability for the drift term $v_k(z)$ to become large
has to be suppressed strongly enough.
More precisely the histogram of the magnitude of the 
drift term should fall off exponentially or faster.
This criterion tells us whether the results obtained by the CLM
are reliable or not.
%% As we mentioned in the previous subsection,
%% the situation in which the CLM fails can be classified into two cases.
%% One is the case in which the complexified variables make long excursions
%% in the imaginary directions 
%% (the excursion problem) \cite{Aarts:2009uq,Aarts:2011ax},
%% and the other is the case in which 
%% the drift term has singularities and the complexified variables come 
%% close to these points frequently 
%% (the singular-drift problem) \cite{Nishimura:2015pba}.
%% In both these cases,
%% the magnitude of the drift term tends to become large, 
%% and the probability distribution of the drift term can have 
%% a power-law behavior at large magnitude.
%% Thus, our criterion can detect these two problems in a unified manner,
%% and more importantly, it enables us to determine precisely
%% the parameter region
%% in which 
%% %the CLM works.
%% these problems occur.
%The usefulness of our criterion 
Its usefulness
was demonstrated in 
ref.~\cite{Nagata:2016vkn}
for two simple one-variable models
and also for systems with many degrees of freedom \cite{Nagata:2018net}.

%\section{Applying the CLM to the Lorentzian model}
%\label{sec:CLM}
\subsection{Applying the CLM to the Lorentzian model}
\label{sec:applying-CLM}

Let us 
%We 
apply the CLM to the model (\ref{Z-Likkt-n1-cutoff}).
From now on, we omit the Pfaffian
%fermionic matrices 
and consider the 
6D version, which consists of $A_0$
% given by (\ref{eq:diagonal gauge})
and $A_i$ ($i=1,\cdots , 5$), for simplicity.

The first step of the CLM is to complexify the real variables.
As for the spatial matrices $A_i$, 
we simply treat them
as general complex matrices instead of Hermitian matrices.
As for the temporal matrix $A_0$, which is diagonalized as
(\ref{eq:diagonal gauge}),
we have to take into account the ordering of the eigenvalues.
For that purpose,
we make the change of variables as
\begin{alignat}{3}
\alpha_1 = 0 \ , \quad
\alpha_2 = \ee^{\tau_1} \ , \quad
\alpha_3 = \ee^{\tau_1} + \ee^{\tau_2} \ ,
\quad
\cdots  \ , \quad 
\alpha_N = \sum_{a=1}^{N-1} 
\ee^{\tau_a}  
  \label{eq:alpha-tau}
\end{alignat}
so that the ordering is implemented automatically,
and then complexify $\tau_a$ ($a=1 ,\cdots , N-1$).
We have chosen to set $\alpha_1 =0$ using the shift symmetry
$A_0 \mapsto A_0 + {\rm const}. {\bf 1}$ of the action.
In order to respect this symmetry,
we 
%decide to 
impose the cutoff like (\ref{eq:t_cutoff})
only on the traceless part 
$\tilde{A}_0 = A_0 - \frac{1}{N} \Tr A_0$
in what follows.
%in this work. 

The Heaviside function in (\ref{Z-Likkt-n1-cutoff}) 
is difficult to treat in the CLM as it is.
Here we mimic its effect by introducing the potential
\begin{alignat}{3}
S_{\rm pot}
= \frac{1}{p} \gamma_{\rm s} 
\left(\frac{1}{N} \Tr (A_i)^2 - 1 \right)^p 
+ \frac{1}{p} \gamma_{\rm t} 
\left(\frac{1}{N} \Tr (\tilde{A}_0)^2 
- \kappa  \right)^p  \ ,
  \label{eq:pot}
\end{alignat}
where the power $p$ is set to $p=4$,
% in this work,
and the coefficients $\gamma_{\rm s} $ and $\gamma_{\rm t}$
are chosen to be large enough to make
$\frac{1}{N} \Tr (A_i)^2$ and $\frac{1}{N} \Tr (\tilde{A}_0)^2$
fluctuate around some constants.\footnote{This appears different
from imposing the inequalities (\ref{eq:t_cutoff}) and
(\ref{eq:s_cutoff}), but
the difference is not important since the inequalities are typically
saturated due to entropic effects.}
The effective action then reads
\begin{alignat}{3}
S_{\rm eff}  
&=  N
\beta
 \,  e^{-i \frac{\pi}{2} (1-s)}  \left\{ 
% &\sim &
\frac{1}{2} e^{-i k \pi}
\Tr [A_0 , A_i]^2 
%% \nonumber \\  &~&
 - \frac{1}{4} 
\Tr [A_i , A_j]^2
\right\} \nonumber \\
& + \frac{1}{p} \gamma_{\rm s} 
\left(\frac{1}{N} \Tr (A_i)^2 - 1 \right)^p 
+ \frac{1}{p} \gamma_{\rm t} 
\left(\frac{1}{N} \Tr (\tilde{A}_0)^2 
- \kappa  \right)^p \nonumber \\
&  - \log \Delta(\alpha)  
- \sum_{a=1}^{N-1}\tau_a \ ,
\label{eq:eff_action}
\end{alignat}
where the last term comes from the Jacobian associated
with the change of variables (\ref{eq:alpha-tau}).
The complex Langevin equation is given by
%can be obtained as
\begin{alignat}{3}
\frac{d\tau_a}{dt}&=
 - \frac{\del S_{\rm eff}}{\del \tau_a} + \eta_a(t)  \ ,   \nn \\
\frac{d (A_i)_{ab}}{dt}&=
 - \frac{\del S_{\rm eff}}{\del (A_i)_{ba}} 
+ (\eta_i)_{ab}(t)  \ ,
  \label{eq:cle}
\end{alignat}
where the $\eta_a(t)$ in the first equation 
are random real numbers
obeying the probability distribution
$\exp (- \frac{1}{4}\int dt \sum_a \{ \eta_a(t) \}^2 )$
and the $\eta_i(t)$ in the second equation 
are random Hermitian matrices
obeying the probability distributions 
$\exp (- \frac{1}{4}\int dt \sum_{i} \Tr  \{ \eta_i(t) \}^2 )$.
%% Let us write $(\eta_i)_{ab}(t)$ in the second equation as
%% $(\eta_i)_{ab}(t)=\sum_A \xi_i^A(t) (T_A)_{ab}$ 
%% with $T_A$ being the SU($N$) generators
%% normalized by $\tr (T_A T_B) = \delta_{AB}$.
%% Then the $\eta_a(t)$ in the first equation and the $\xi_i^A(t)$ 
%% introduced above are real Gaussian random variables
%% obeying the probability distributions 
%% $\exp (- \frac{1}{4}\int dt \sum_a \{ \eta_a(t) \}^2 )$ and
%% $\exp (- \frac{1}{4}\int dt \sum_{iA} \{ \xi_i^A(t) \}^2 )$,
%% respectively.

The expectation values of observables can be calculated by 
defining them holomorphically for complexified $\tau_a$ and $A_i$
and taking an average using the configurations generated by
solving the discretized version of (\ref{eq:cle}) 
for sufficiently long time.
In order for this method to work, the probability distribution
of the drift terms, namely the first terms
on the right-hand side of (\ref{eq:cle}), has to fall off 
exponentially \cite{Nagata:2016vkn}.
We have checked that this criterion is indeed satisfied
for all the values of parameters used in this paper.

%%%%%%%%%%%%%%%%%%%%%%%%%%%%%%%%%%%%%%%%%%%%%%%%%%%%%%%%%%%%%%%%%%%%%%
%%%%%%%%%%%%%%%%%%%%%%%%%%%%%%%%%%%%%%%%%%%%%%%%%%%%%%%%%%%%%%%%%%%%%%
%%%%%%%%%%%%%%%%%%%%%%%%%%%%%%%%%%%%%%%%%%%%%%%%%%%%%%%%%%%%%%%%%%%%%%\
\section{Emergence of (3+1)D expanding behavior}
\label{sec:expanding}
%%%%%%%%%%%%%%%%%%%%%%%%%%%%%%%%%%%%%%%%%%%%%%%%%%%%%%%%%%%%%%%%%%%%%%
%%%%%%%%%%%%%%%%%%%%%%%%%%%%%%%%%%%%%%%%%%%%%%%%%%%%%%%%%%%%%%%%%%%%%%
%%%%%%%%%%%%%%%%%%%%%%%%%%%%%%%%%%%%%%%%%%%%%%%%%%%%%%%%%%%%%%%%%%%%%%\

In this section we consider $(s,k)=(-1,0)$
in the parameter space.
The action is given by
\begin{alignat}{3}
S = N \beta \,
\Big\{ 
- \frac{1}{2}  \Tr [A_0 , A_i]^2 
+ \frac{1}{4} \Tr [A_i , A_j]^2 
\Big\}  \ ,
\label{sdef-action3}
\end{alignat}
which is real, and the CLM reduces 
to the ordinary Langevin method.
The first term in (\ref{sdef-action3})
tries to minimize the space-time noncommutativity,
which has the effects of making
the spatial matrices close to diagonal
in the basis (\ref{eq:diagonal gauge}).
On the other hand, the second term favors maximal noncommutativity
among spatial matrices.

\begin{figure}[t]
\centering
\includegraphics[width=7cm]{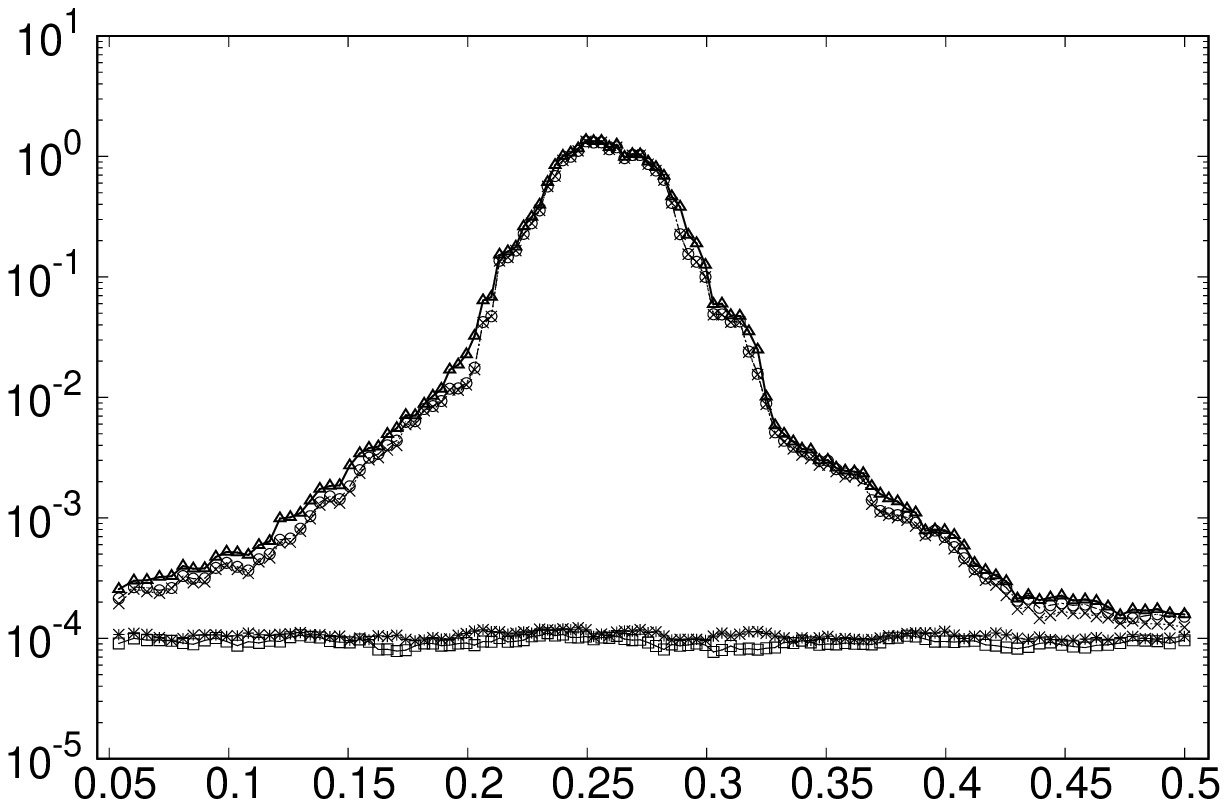}
\includegraphics[width=7cm]{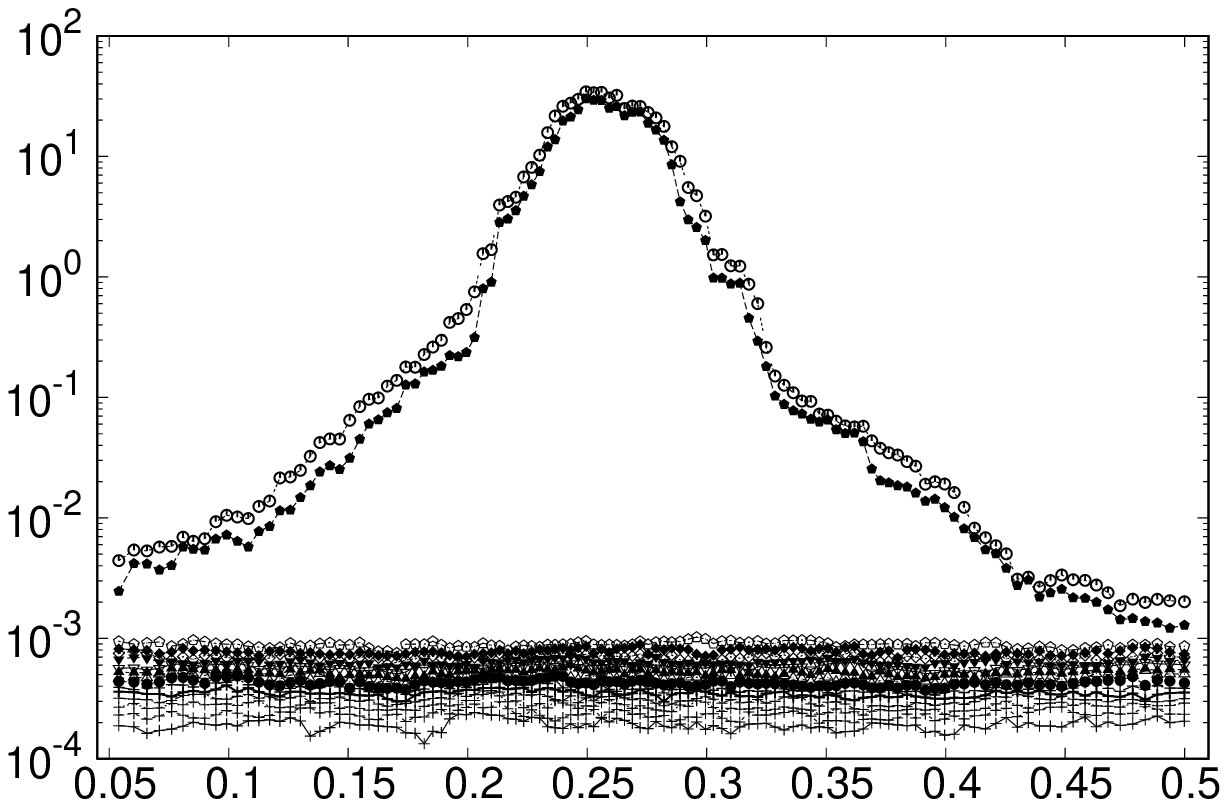}
\caption{Results for $(s,k)=(-1,0)$, $N=128$, $\kappa=0.02$, $\beta=8$,
$n=16$ are shown.
%% (Top) The extent of space $R^2(t)$ is plotted
%% against $t$.
%(Bottom-Left) 
(Left) 
The five eigenvalues of the moment of inertia tensor
are plotted against $t$ in the log scale.
%(Bottom-Right) 
(Right) 
The 16 eigenvalues of the matrix $Q(t)$ are plotted
against $t$ in the log scale.}
%% For $\alpha=0.2$, one finds many eigenvalues distributing near the origin,
%% while for $\alpha=0.4$, there is a gap near the origin.}
\label{fig:N128sdef-1}
\end{figure}

Figure \ref{fig:N128sdef-1} 
shows our results\footnote{Here and hence forth,
we plot the results obtained for one thermalized configuration.}
for $N=128$, $\kappa=0.02$, $\beta=8$.
The block size for (\ref{eq:def_abar}) is chosen to be $n=16$.
%% In the Top panel, we plot the 
%% extent of space $R^2(t)$ defined
%% by (\ref{eq:def_rsq}) against $t$.
%% The result is symmetric under the reflection 
%% $t - t_{\rm p} \mapsto -(t - t_{\rm p})$, where
%% $t_{\rm p}$ represents the time at which $R^2(t)$ is peaked,
%% due to the symmetry of the model 
%% under $\tilde{A}_0 \mapsto - \tilde{A}_0$.
%
%Next we discuss the SSB of SO(5) symmetry by
%considering the moment of inertia tensor (\ref{eq:def_tij}).
%The block size is chosen to be $n=16$.
In the Left panel,
%In the Bottom-Left panel,
%Fig.~\ref{fig:N=128sdef=-1} (Bottom-Left) 
we plot the 
eigenvalues $\lambda_{i}(t)$ of $T_{ij}(t)$, which
shows that only three out of five eigenvalues become large
in the time region around 
the peak.
%$t=t_{\rm p}$.
This suggests that the rotational SO(5) symmetry 
of the 6D bosonic model is broken down to SO(3) in that time region.
These results are qualitatively the same as
what has been obtained in ref.~\cite{Ito:2015mxa},
which is consistent with the speculation \cite{Aoki:2019tby}
that the previous simulations correspond to 
the parameter choice $(s,k)=(-1,0)$.
% instead of $(s,k)=(0,0)$ corresponding to the original model.

As is known from the previous work \cite{Kim:2011cr},
the time difference between the peak 
%($t=t_{\rm p}$)
and the critical time at which the SSB occurs
increases in physical units
as we take the large-$N$ limit.
Therefore, the reflection symmetry with respect to $t$
does not necessarily
imply that the Big Crunch occurs in the finite future.

The mechanism of this SSB can be understood as follows \cite{Aoki:2019tby}.
Since the first term in (\ref{sdef-action3}) favors $A_i$ close to diagonal, 
we may consider the submatrices $\bar{A}_i(t)$ as the effective
degrees of freedom.
The infrared cutoff (\ref{eq:s_cutoff}) fixes
$\Tr \{ \bar{A}_i(t) \} ^2$ to some constant, and 
the second term in (\ref{sdef-action3}) favors maximal noncommutativity
between $\bar{A}_i(t)$.
According to the argument in ref.~\cite{Kim:2011cr}, this leads
to $\bar{A}_i(t)\propto \sigma_i \oplus {\bf 0}_{n-2}$ 
for $i=1,2,3$ and 
$\bar{A}_i(t) = {\bf 0}_{n}$ for $i\ge 4$ up to SO(5) rotations,
where $\sigma_i$ are the Pauli matrices.
In order to confirm this mechanism, we calculate the matrix 
\begin{alignat}{3}
Q(t)=\sum_{i=1}^5 \{ \bar{A}_i (t) \}^2 \ ,
\label{def-Q}
\end{alignat}
and plot the 
%$Q(t)=\sum_{i=1}^5 \{ \bar{A}_i (t) \}^2$ and plot the 
%four largest 
eigenvalues of $Q(t)$ in 
%Fig.~\ref{fig:N128sdef-1} (Bottom-Right).
Fig.~\ref{fig:N128sdef-1} (Right).
%the Bottom-Right panel.
Indeed we find 
that only two of them are large, while the rest are very small
in the time region in which the SSB occurs.

%[jnishi@momo kappa=0.05]$ pwd
%/home/jnishi/collab/codes/cle-ikkt/susy6d/c_kappa/action_kdef/N=128/kappa=0.05
%[jnishi@momo kappa=0.05]$ 

\begin{figure}[t]
\centering
\includegraphics[width=7cm]{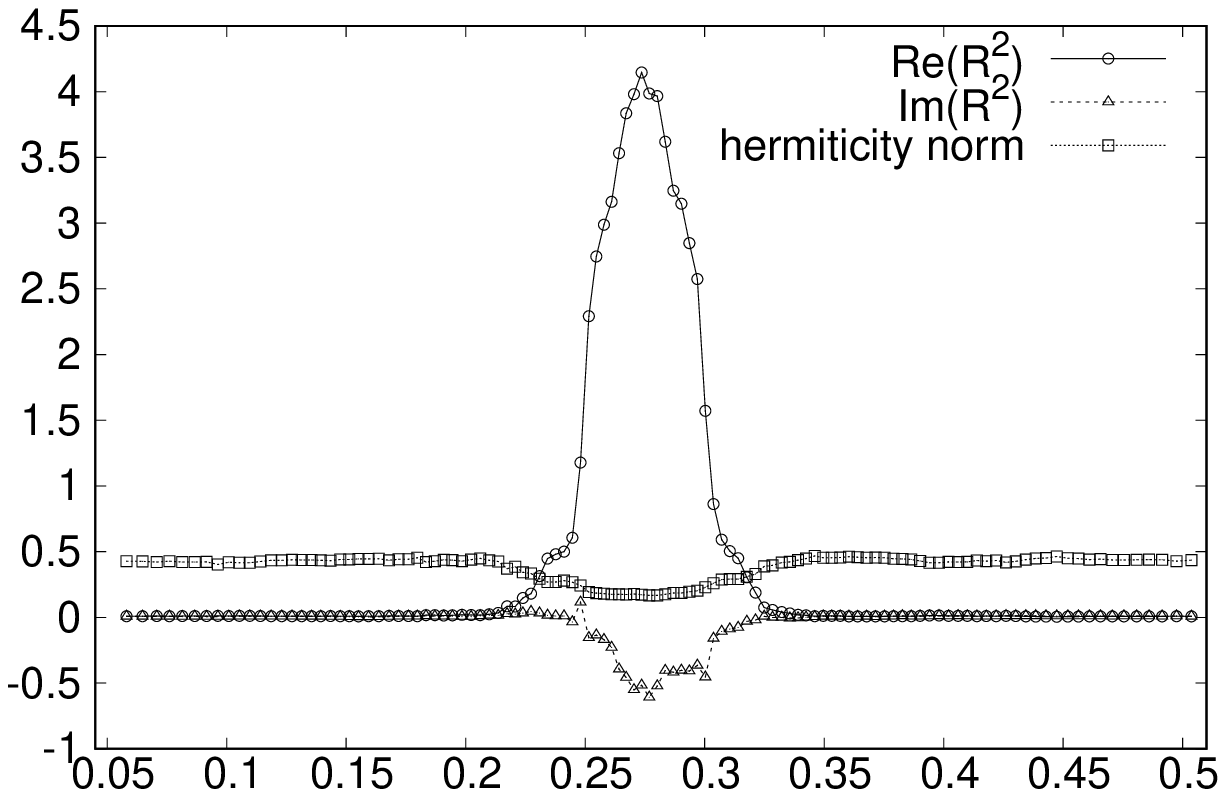}\\
\includegraphics[width=7cm]{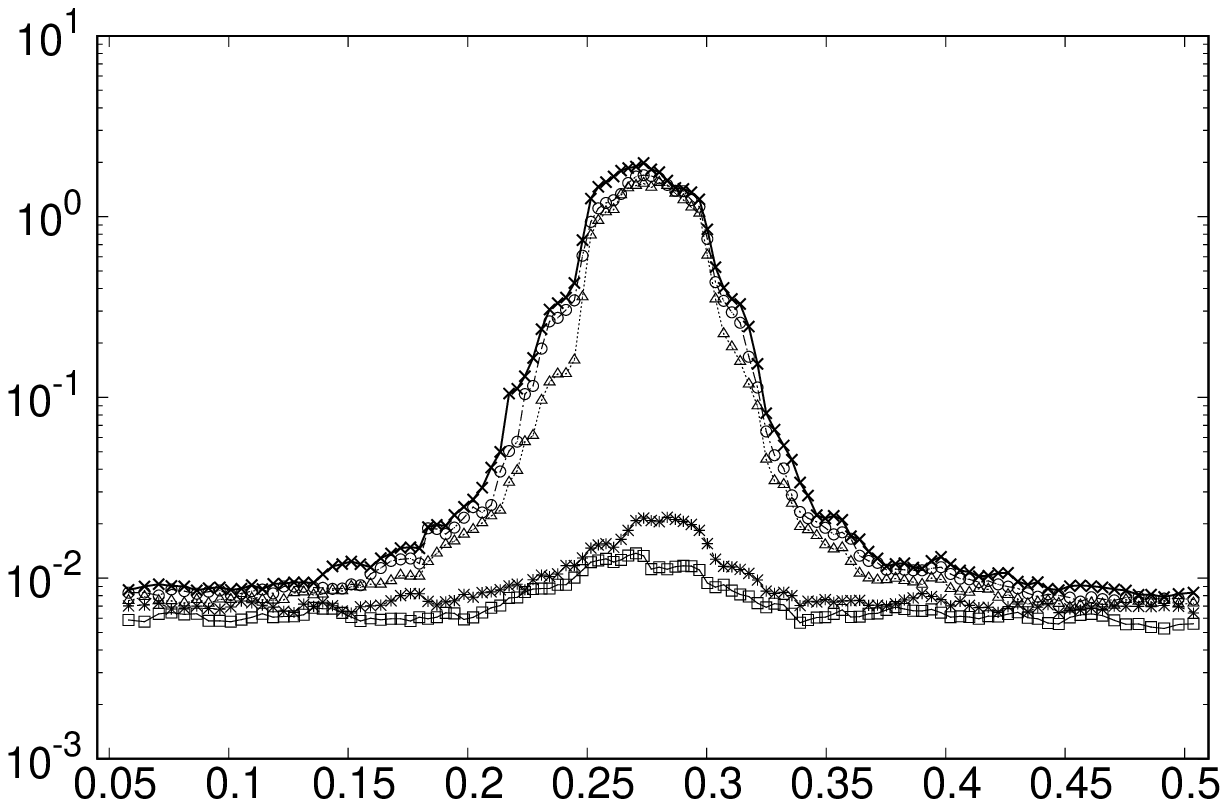}
\includegraphics[width=7cm]{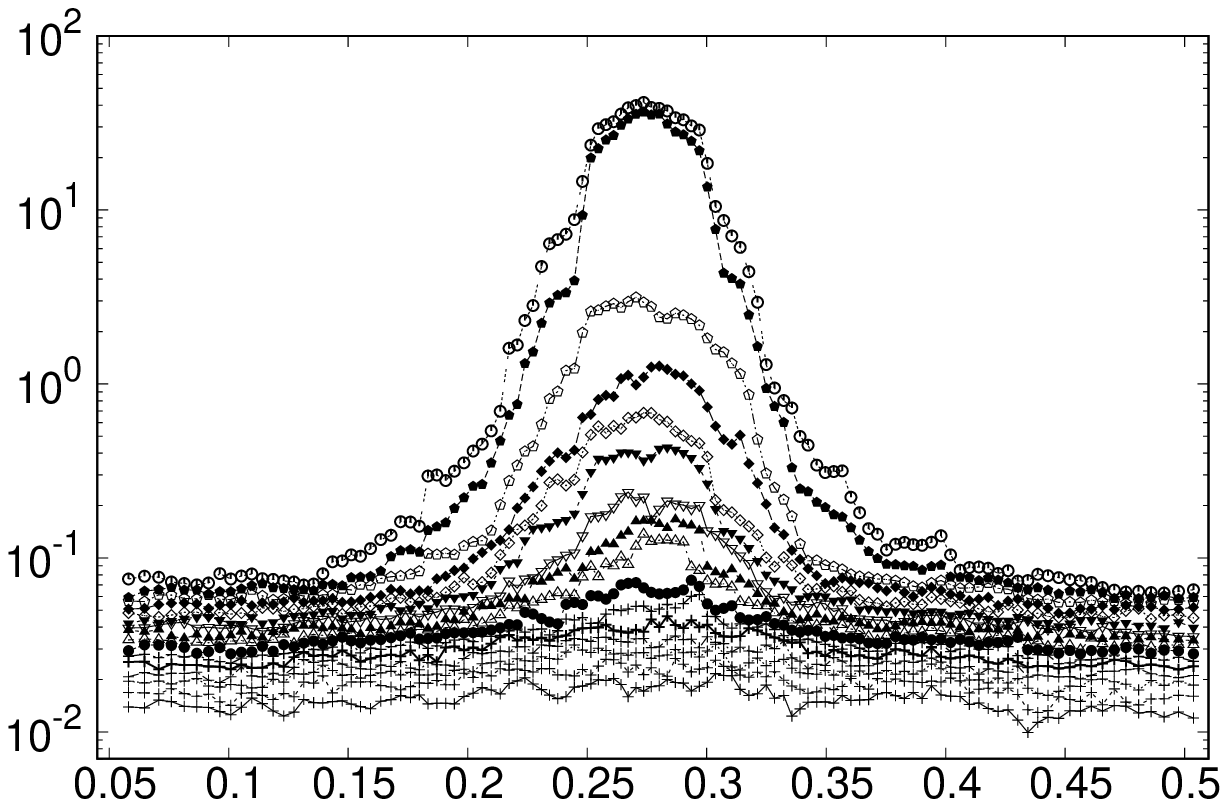}

%\\[0mm]
\caption{Results for $(s,k)=(-0.004,0.498)$, 
$N=128$, $\kappa=0.02$, $\beta=8$, $n=16$ are shown.
%% \caption{Results for $(s,k)=(0.0076,0.5038)$, 
%% $N=128$, $\kappa=0.0037$, $\beta=32$, $n=16$ are shown.
(Top) The real and imaginary parts of $R^2(t)$ are plotted
against $t$. 
% by plus symbols and cross symbols, respectively.
The Hermiticity norm $h(t)$ of the matrix $\bar{A}_i(t)$ is also plotted.
%by the asterisk symbols.
(Bottom-Left) The five eigenvalues of the moment of inertia tensor
are plotted against $t$ in the log scale.
(Bottom-Right) The 16 eigenvalues of the matrix $Q(t)$ are plotted
against $t$ in the log scale.
}
%% For $\alpha=0.2$, one finds many eigenvalues distributing near the origin,
%% while for $\alpha=0.4$, there is a gap near the origin.}
\label{fig:N128sdef0.0076}
\end{figure}

%%%%%%%%%%%%%%%%%%%%%%%%%%%%%%%%%%%%%%%%%%%%%%%%%%%%%%%%%%%%%%%%%%%%%%
%%%%%%%%%%%%%%%%%%%%%%%%%%%%%%%%%%%%%%%%%%%%%%%%%%%%%%%%%%%%%%%%%%%%%%
%%%%%%%%%%%%%%%%%%%%%%%%%%%%%%%%%%%%%%%%%%%%%%%%%%%%%%%%%%%%%%%%%%%%%%\
%\section{Departure from the Pauli-matrix structure}
\section{Emergence of a smooth space-time}
\label{sec:departure}
%%%%%%%%%%%%%%%%%%%%%%%%%%%%%%%%%%%%%%%%%%%%%%%%%%%%%%%%%%%%%%%%%%%%%%
%%%%%%%%%%%%%%%%%%%%%%%%%%%%%%%%%%%%%%%%%%%%%%%%%%%%%%%%%%%%%%%%%%%%%%
%%%%%%%%%%%%%%%%%%%%%%%%%%%%%%%%%%%%%%%%%%%%%%%%%%%%%%%%%%%%%%%%%%%%%%\

In this section we tune the worldsheet deformation parameter $s$ to
some values near $s=0$, 
which is the target value for the Lorentzian model, 
keeping the target-space deformation parameter $k$ to be
$k=(1+s)/2$, which minimizes the space-time noncommutativity.
The action reads
\begin{alignat}{3}
S = N \beta \,
\Big\{ - \frac{1}{2}  \Tr [A_0 , A_i]^2 
- \ee^{-i\frac{\pi}{2}(1-s)}
\frac{1}{4} \Tr [A_i , A_j]^2 
\Big\}  \ .
\label{sdef-action4}
\end{alignat}
The only difference from (\ref{sdef-action3})
is the second term
with the coefficient $\ee^{-i\frac{\pi}{2}(1-s)}$
whose real part changes its sign at $s=0$.
This implies, in particular, that for $s>0$ 
the second term starts to minimize the noncommutativity
among the spatial matrices.
Therefore, we may anticipate a drastic change of the behavior around $s=0$.
In fact, for the values of $s$ below what is reported below, we
do not see any qualitative difference from the results obtained at $s=-1$.

Figure \ref{fig:N128sdef0.0076}
%Fig.~\ref{fig:N=128r2-sdef0.0076} 
shows our results
for $N=128$, $\kappa=0.02$, $\beta=8$, $n=16$
with $(s,k)=(-0.004,0.498)$.
%with $(s,k)=(0.0076,0.5038)$.
Unlike the $(s,k)=(-1,0)$ case, 
the action becomes complex 
for $s>-1$ in general.
%in the present case.
Therefore, the quantity such as $R^2(t)$ defined in (\ref{eq:def_rsq})
is not guaranteed to be real positive.
In the Top panel,
we plot the real and imaginary parts of $R^2(t)$.
%% Here we evaluate $R^2(t)$ by calculating
%% $\frac{1}{n}{\rm tr}\sum_{i}
%% \left(\bar{A}_{i}\left(t\right)\right)^{2}$
%% for one complex 
%% %$\bar{A}_{i}(t)$ obtained from 
%% configurations generated by the complex Langevin simulation.
%without taking the Hermitian part.
We find that $R^2(t)$ is dominated by the real part
near the peak.

Let us also take a look at the ``Hermiticity norm''
for $\bar{A}_{i}(t)$ defined by
\begin{alignat}{3}
h(t) =
\frac{- \Tr 
(\bar{A}_i(t) - \bar{A}_i(t)^\dag)^2}
{4 \, \Tr (\bar{A}_i(t) ^\dag \bar{A}_i(t))}
\label{hermiticity-norm} \ ,
\end{alignat}
using a configuration generated by the simulation.
The result is plotted in the 
%Fig.~\ref{fig:N=128sdef=0.0076} (Top)
Top panel
as well.
Note that $h(t)=0$ implies
that the matrices $\bar{A}_i(t)$ are all Hermitian,
while $h(t)=1$ implies
that they are all anti-Hermitian.
We find that $h(t)$ is small and hence
the $\bar{A}_{i}(t)$ are close to Hermitian
near the peak,
which is consistent with 
our observation that $R^2(t)$ is dominated by the real part
in this region.
%First of all, this is a good news because otherwise we run into a
%strange situation that the space is not ``real''.
This property supports our previous 
speculation \cite{Kim:2011ts,Kim:2012mw}
that some classical solution, which is typically 
represented by a real configuration,
dominates the path integral in the time region near the peak
due to the expansion of space.

%In Fig.~\ref{fig:N=128sdef=0.0076} (Bottom) 
In the Bottom panels, 
we plot the 
same quantities\footnote{In fact,
it is not straightforward to calculate the expectation values of
the eigenvalues of $T_{ij}(t)$ and $Q(t)$ in the CLM
respecting holomorphicity because of their multi-valuedness.
Here we simply evaluate $T_{ij}(t)$ and $Q(t)$
using the Hermitian part of $\bar{A}_{i}(t)$
from one configuration
generated by the complex Langevin simulation,
and plot their eigenvalues.}
%%
%% In fact,
%% it is not straightforward to calculate the expectation values of
%% the eigenvalues of $T_{ij}(t)$ and $Q(t)$ in the CLM
%% because of their multi-valuedness.
%%
as in Fig.~\ref{fig:N128sdef-1}.
% (Bottom).
% for $\kappa=0.0037$, $\beta=32$, $s=0.0076$. 
%The block size is chosen to be $n=16$ as before.
From the left panel, we observe that (3+1)D expanding behavior
persists at $s \sim 0$, while the right panel reveals
a clear departure from the Pauli-matrix structure.

We have also investigated the model with $N<128$.
For $N=32$, $64$, while the results at $(s,k)=(-1,0)$ are similar
to those for $N=128$,
the departure from the Pauli-matrix structure does not show up 
at all even at $s \sim 0$.
As we increase $s$ further with $k=(1+s)/2$,
the Hermiticity
of the configurations is completely lost, and the criterion for 
justifying the CLM is found to be violated.
Thus the use of large values of $N$ seems to be crucial
in investigating the model near the target values $(s,k)=(0,0)$.
%ing a smooth space-time picture in this model.

%/home/jnishi/collab/codes/cle-ikkt/susy6d/c_kappa/action_kdef/N=192/kappa=0.001/beta=64/sdef=0.0118

%%%%%%%%%%%%%%%%%%%%%%%%%%%%%%%%%%%%%%%%%%%%%%%%%%%%%%%%%%%%%%%%%%%%%%
%%%%%%%%%%%%%%%%%%%%%%%%%%%%%%%%%%%%%%%%%%%%%%%%%%%%%%%%%%%%%%%%%%%%%%
%%%%%%%%%%%%%%%%%%%%%%%%%%%%%%%%%%%%%%%%%%%%%%%%%%%%%%%%%%%%%%%%%%%%%%\
\section{Summary and discussions}
\label{sec:summary}
%%%%%%%%%%%%%%%%%%%%%%%%%%%%%%%%%%%%%%%%%%%%%%%%%%%%%%%%%%%%%%%%%%%%%%
%%%%%%%%%%%%%%%%%%%%%%%%%%%%%%%%%%%%%%%%%%%%%%%%%%%%%%%%%%%%%%%%%%%%%%
%%%%%%%%%%%%%%%%%%%%%%%%%%%%%%%%%%%%%%%%%%%%%%%%%%%%%%%%%%%%%%%%%%%%%%\

%% In this paper we have discussed the space-time structure
%% in the Lorentzian type IIB matrix model.
We consider that by now there is surmounting evidence that
the Lorentzian type IIB matrix model is a promising candidate
for a nonperturbative formulation of superstring theory.
Unfortunately Monte Carlo studies of this model are extremely hard
due to the sign problem caused by 
the phase factor $\ee^{iS_{\rm b}}$ in the partition function.
Previous work avoided this problem by integrating out the scale factor
of the bosonic matrices and using an approximation.
However, it was noticed recently that
this approximation actually amounts to 
replacing $\ee^{iS_{\rm b}}$ by $\ee^{c S_{\rm b}}$ for some $c>0$.
This suggests the importance of studying the model 
without such an approximation.

In this article we have reviewed the results of 
ref.~\cite{Nishimura:2019qal},
where we investigated the space-time structure
based on the complex Langevin simulation 
of the (5+1)D bosonic version of the model with the deformation parameters
$s$ and $k$ corresponding to the Wick rotations on the worldsheet and 
in the target space, respectively,
%The action is given by (\ref{sdef-action2}).
%
The original model corresponds to $(s,k)=(0,0)$,
whereas our previous simulations were speculated to
correspond to the $(s,k)=(-1,0)$ case \cite{Aoki:2019tby}.
%% The original model, which corresponds to $(s,k)=(0,0)$, is
%% extremely difficult to study because the partition function involves
%% the phase factor $\ee^{iS_{\rm b}}$
Our results for $(s,k)=(-1,0)$ indeed
reproduced the (3+1)D expanding behavior with 
the Pauli-matrix structure as expected.
Then we tuned the parameter $s$ towards 
the region $s \sim 0$
restricting ourselves to $k=(1+s)/2$ in order to stabilize
our simulation.
%%  although we expect that the parameter region
%% close to the original model $(s,k)=(0,0)$ becomes accessible 
%% at larger $N$.
The results indeed showed a clear departure from
the Pauli-matrix structure, while the (3+1)D expanding behavior
is kept intact.
The spatial matrices turn out to be close to Hermitian near the peak 
of the spatial extent $R^2(t)$ even for $s\sim 0$,
which confirms our expectation \cite{Kim:2011ts,Kim:2012mw}
that some classical solution dominates
at late times.

The appearance of the Pauli-matrix structure
for $(s,k)=(-1,0)$
is due to the $\Tr (F_{ij})^2$ term in the action, which tries
to make the spatial matrices $A_i$ maximally noncommutative.
The situation changes drastically around $s=0$, where
the coefficient of the $\Tr (F_{ij})^2$ term becomes
pure imaginary.
On the other hand, there are infinitely many classical 
solutions \cite{Kim:2011ts,Kim:2012mw,Hatakeyama:2019jyw,Klinkhamer:2019lhp},
which have (3+1)D expanding behavior without
the Pauli-matrix structure.
(See also refs.\cite{Chaney:2015ktw,Chaney:2015mfa,Chaney:2016npa,Stern:2018wud,Steinacker:2017vqw,Steinacker:2017bhb} for related work.)
We therefore consider it possible that
the space-time structure becomes smooth
without losing the (3+1)D expanding behavior 
in the large-$N$ limit.
The departure from the Pauli-matrix structure
observed at $s\sim 0$ supports this possibility.

As future prospects, 
the most important thing to do is to repeat the same analysis
with increased matrix size $N$.
In particular, we need to confirm the appearance of 
a smooth space-time at $(s,k)\sim (0,0)$.
While this issue may not depend much on 
the effects of the fermionic matrices,
it would be certainly desirable to include them eventually.
Unfortunately, this is not straightforward
since the complex Langevin method may suffer from the singular-drift
problem due to the near-zero eigenvalues of the Dirac operator.
The deformation technique \cite{Ito:2016efb}
used successfully in studying
the Euclidean version \cite{Anagnostopoulos:2017gos,Anagnostopoulos:2020xai}
is worth trying, though.
We consider that the dominance of classical solutions at late times
\cite{Kim:2011ts,Kim:2012mw}
supported by our results is important because it enables us to understand
possible late-time behaviors of this model
by solving classical equations of motion.
%We have developed a code for generating classical solutions numerically
For instance, we may try to find classical 
solutions \cite{Chatzistavrakidis:2011gs,Nishimura:2013moa,Steinacker:2014fja,Aoki:2014cya},
which can accommodate Standard Model particles as excitations around them.
See ref.~\cite{Hatakeyama:2019jyw} for a recent work in this direction,
where we also find that smooth space-time structure is obtained
quite generically in classical solutions.

The new perspectives 
on the (3+1)D expanding space-time in the Lorentzian type IIB 
matrix model that have arised recently make this model more
promising as a nonperturbative formulation of superstring theory.
In particular, it is encouraging that the CLM turns out to be useful
in overcoming the severe sign problem that occurs in applying
standard Monte Carlo methods to this model.
On the other hand, further developments in solving the sign problem
might be necessary in order to approach the target theory
corresponding to $(s,k)=(0,0)$ including the fermionic matrices.
We hope that such developments eventually enable us to answer
profound questions concerning the origin of our Universe.

%%%%%%%%%%%%%%%%%%%%%%%%%%%%%%%%%%%%%%%%%%%%%%%%%%%%%%%%%%%%%%%%%%%%%%
%%%%%%%%%%%%%%%%%%%%%%%%%%%%%%%%%%%%%%%%%%%%%%%%%%%%%%%%%%%%%%%%%%%%%% 

%%%%%%%%%%%%%%%%%%%%%%%%%%%%%%%%%%%%%%%%%%%%%%%%%%%%%%%%%%%%%%%%%%%%%%
%%%%%%%%%%%%%%%%%%%%%%%%%%%%%%%%%%%%%%%%%%%%%%%%%%%%%%%%%%%%%%%%%%%%%%
%%%%%%%%%%%%%%%%%%%%%%%%%%%%%%%%%%%%%%%%%%%%%%%%%%%%%%%%%%%%%%%%%%%%%%
\section*{Acknowledgements}
%\hspace{0.51cm}
%%%%%%%%%%%%%%%%%%%%%%%%%%%%%%%%%%%%%%%%%%%%%%%%%%%%%%%%%%%%%%%%%%%%%%
%%%%%%%%%%%%%%%%%%%%%%%%%%%%%%%%%%%%%%%%%%%%%%%%%%%%%%%%%%%%%%%%%%%%%%
%%%%%%%%%%%%%%%%%%%%%%%%%%%%%%%%%%%%%%%%%%%%%%%%%%%%%%%%%%%%%%%%%%%%%% 

The author would like to thank 
K.N.~Anagnostopoulos,
T.~Aoki, T.~Azuma,
M.~Hirasawa, Y.~Ito,
H.~Kawai, 
S.K.~Papadoudis, 
H.~Steinacker and A.~Tsuchiya for valuable discussions.
He is also grateful to 
K.N.~Anagnostopoulos and G.~Zoupanos for their warm hospitality 
during the Summer Institute.
%%This work was supported by MEXT as "Program for Promoting 
%%Researches on the Supercomputer Fugaku" 
This work was supported by MEXT as ``Program for Promoting 
Researches on the Supercomputer Fugaku''
(Simulation for basic science: from fundamental laws of particles
to creation of nuclei).
It was also supported by computational time granted from the Greek
Research \& Technology Network (GRNET) in the 
National HPC facility -ARIS- under project IDs IIB10D and LIKKT.

%conference.
%% Computation was carried out
%% on PC clusters at KEK and XC40 at YITP in Kyoto University.
%% J.~N.\ and A.~T.\ were supported in part by Grant-in-Aid 
%% for Scientific Research (No.\ 16H03988 and 18K03614, respectively)
%% from Japan Society for the Promotion of Science. 
%%
%% This work was supported by computational time granted from the Greek
%% Research \& Technology Network (GRNET) in the 
%% National HPC facility -ARIS- under project ID LIKKT.

%% \begin{thebibliography}{99}
%% \bibitem{...}
%% ....
%% \end{thebibliography}

\bibliographystyle{JHEP}
\bibliography{cle-lorentz_ref}

\end{document}